\documentclass[review]{elsarticle}

\usepackage{graphicx}
\DeclareGraphicsExtensions{.eps}

\usepackage{color}

\newcommand{\denselist}{\setlength{\itemsep}{0cm} \setlength{\parskip}{0cm}}

\begin{document}

\begin{frontmatter}


\title{A feasibility study to track cosmic muons using a detector with SiPM devices based on amplitude discrimination}







\author[label1]{D.~Stanca\corref{cor1}}
\ead{denis.stanca@gmail.com}
\author[label1]{M.~Niculescu-Oglinzanu\corref{cor1}}
\ead{mihai.niculescu@nipne.ro}
\author[label1]{I.~Brancus}
\author[label1]{B.~Mitrica}
\author[label1]{A.~Balaceanu}
\author[label2]{B.~Cautisanu}
\author[label1]{A.~Gherghel-Lascu}
\author[label3]{A.~Haungs}
\author[label3]{H.-J.~Mathes}
\author[label3]{H.~Rebel}
\author[label1]{A.~Saftoiu}
\author[label2]{O.~Sima}
\author[label1]{T.~Mosu}

\address[label1]{Horia Hulubei Institute of Physics and Nuclear Engineering (IFIN-HH), Bucharest, P.O.B. MG-6, Romania}
\address[label2]{Department of Physics, University of Bucharest, P.O.B. MG-11, Romania}
\address[label3]{Karlsruhe Institute of Technology, Institut f\"ur Kernphysik, 76021 Karlsruhe, Germany}

\cortext[cor1]{Corresponding authors}


\begin{abstract}
The possibility to build a SiPM-readout muon detector (SiRO), 
using plastic scintillators with optical fibers as sensitive volume and readout by 
SiPM photo-diodes, is investigated. SiRO shall be used for tracking cosmic muons 
based on amplitude discrimination.
The detector concept foresees a stack of 6 active layers, grouped in 3 sandwiches for 
determining the muon trajectories through 3 planes. After investigating the characteristics 
of the photodiodes, tests have been performed using two detection modules, each being composed 
from a plastic scintillator sheet, $100 \times 25 \times 1\,$cm$^{3}$, with 12 parallel, 
equidistant ditches;
each ditch filled with an optical fiber of $1.5\,$mm thickness and always two fibers 
connected to form a channel. 
The attenuation of the light response along the optical fiber and across the 
channels have been tested. The measurements of the incident muons based on the 
input amplitude discrimination indicate that this procedure is not efficient and therefore not 
sufficient, as only about 30\% of the measured events could be used in the reconstruction of 
the muon trajectories.
Based on the studies presented in this paper, the layout used for building the SiRO 
detector will be changed as well as the analog acquisition technique will be replaced 
by a digital one.
\end{abstract}

\begin{keyword}
muon \sep underground \sep SiPM \sep MPPC
\MSC[2010] 00-01\sep  99-00
\end{keyword}

\end{frontmatter}


\section{Introduction}

Silicon-Photomultiplier (SiPM) are semiconductor based photo-sensors that offer several 
advantages compared to other photo detection devices, like the classical photomultipliers:
\begin{itemize}
\denselist
\item higher photon detection probability
\item improved time resolution
\item reduced power consumption pro active area
\item considerable reduced need of volume for installation
\item reduced electric  voltage for operation
\item mechanical robustness
\item insensitivity for  high magnetic fields
\end{itemize}
These features have prompted an interesting technological development for actual and future 
experiments (e.g.~see~\cite{Kovaltchouk, Yebras, Hayato}) being also considered to be 
suited for astroparticle physics experiments, in particular with the aim 
of the observation of particles produced by primary cosmic rays in the atmosphere. 
The features have prompted ideas of dedicated applications as detectors for muons 
(in particular together with optical fibers) in various different fields, also for example in
establishing `muon tomography', i.e.~giving access to the opacity of archaeological and 
geological structures \cite{Aguillar, Marteau}.
In this course we aim to setup a multi-purpose, mobile muon tracking detector based on 
SiPM readout, called SiRO, the SiPM ReadOut muon detector.
Such a tracking detector will first be used underground in the Slanic salt mine, caracterised in \cite{Stanca}, to 
determine muon multiplicities and to analyze the structure of the salt layers. 
The aim of the present paper is to investigate the possibility to determine the 
muon incidence and their trajectories using scintillator layers readout through optical fibers 
viewed by SiPM devices and analysing the amplitude of the events.
The general concept of the SiRO detector is presented as well as
tests and measurements with a setup of two first detection modules are described.

\section{The concept of the SiRO detector layout}

\begin{figure}[h]
\begin{center}
\includegraphics[width=\textwidth]{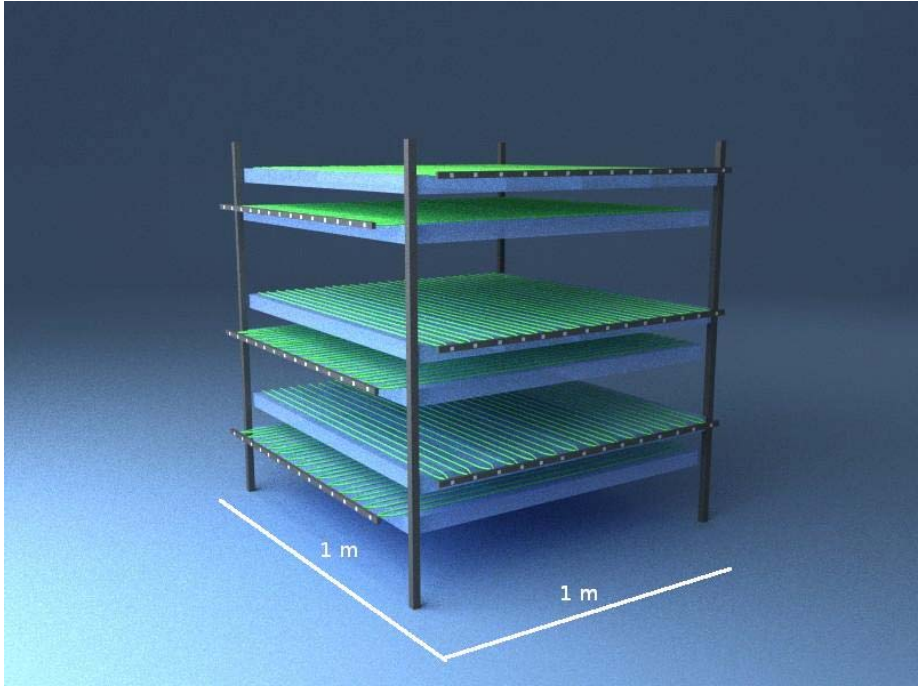}
\caption{\label{fig1} Schematic view of the SiRO detector.}
\end{center}
\end{figure}

\begin{figure}[h]
\begin{center}
\includegraphics[width=8cm, height=10cm]{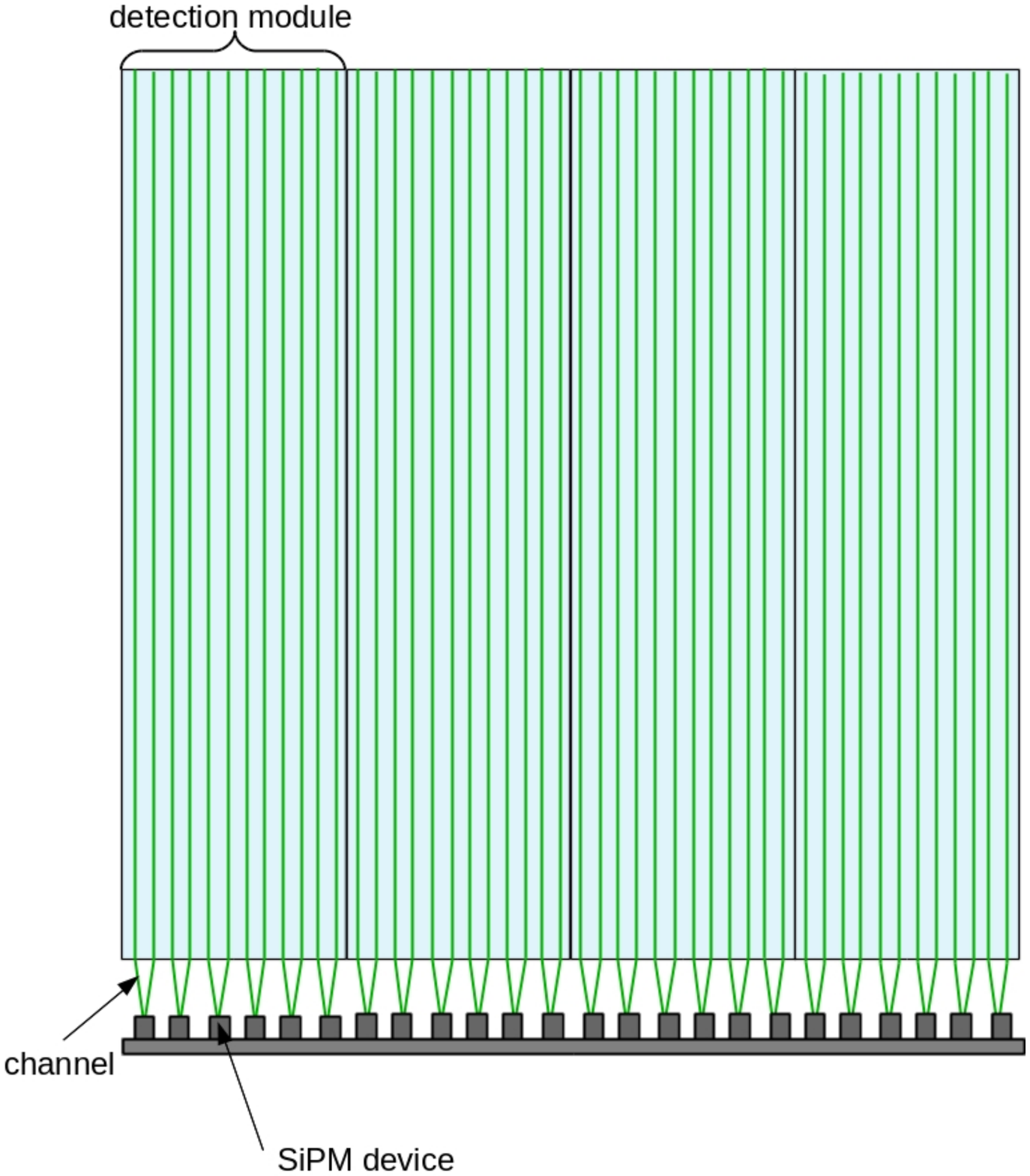}
\caption{\label{fig2} Sketch of an active layer with the definition of one 
detection module and of a single channel.}
\end{center}
\end{figure}

SiRO is designed for flux measurements and arrival direction 
identification~\cite{Brancus} of cosmic muons, and planned to be installed first in an 
underground location, i.e. the Unirea salt mine from Slanic Prahova, Romania. 
The fist prototype is composed of 6 active layers (Fig.~\ref{fig1}), 
each layer consisting of 4 detection modules. 
A detection module uses as sensitive volume a scintillator plate 
(Polystyrol 80 \%, Methylmetacrylate 20 \%) of $100~\times~25~\times~1\,$cm$^3$ 
with 12 parallel and equidistant ditches on its surface, each ditch filled with 
optical fibers. 
Two adjacent optical fibers are connected to a SiPM device to form a channel, 
so that each detection module have six channels.
In Figure~\ref{fig2} a sketch of one active layer is presented.

Each group of two active layers (from top to bottom), with the optical fibers positioned 
on perpendicular directions, represent a sandwich, which should determine the position 
in the input \textit{XY} plane of the incident charged particle. Thus, as we can see in 
Fig.~\ref{fig1}, six active layers with four SiRO modules each, grouped in three 
sandwiches, are put in coincidence to allow the reconstruction of the muon trajectory. 

\begin{figure}[!h]
\begin{center}
\includegraphics[width=\textwidth]{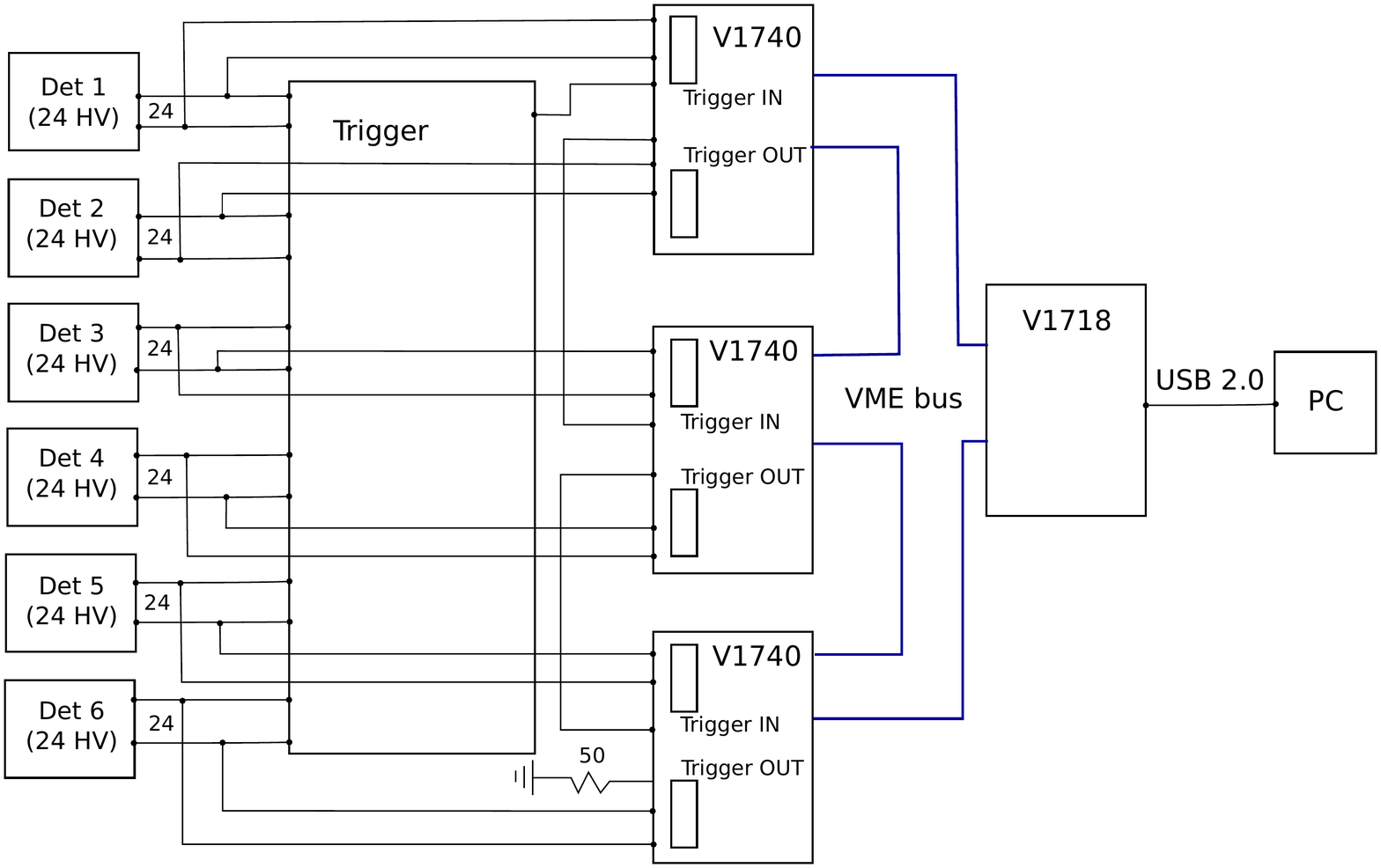}
\caption{\label{fig3} The block scheme of the acquisition system~\cite{Crisan}.}
\end{center}
\end{figure}

Fig.~\ref{fig3} shows the block scheme of the acquisition system~\cite{Crisan}. 
The six active layers are indicated as Det1 to Det6 and contain 
24 channels each, so the whole system will give information from 144 individual channels.

The 24 signals from each active layer are used as an input by a trigger module to create 
coincidences by a certain multiplicity criteria in order to produce a trigger signal, 
marking an event of interest.
The trigger signal is sent by daisy chain along 3 modules 64 Channel Digitizer (V1740 CAEN). 
These modules allow to open an acquisition window with selected pre-trigger time, compensating 
in this way the necessary delay to produce the trigger pulse. 
A USB2.0 Bridge (V1718 CAEN) is used to transfer the digitized signals to a PC.

Each channel of the digitizer has a SRMA memory, where the event can be readout by using the 
PCI-VME bridge module. The events are read sequentially and transmitted to the computer.
The selection of the real events, as well as the track reconstruction is performed off-line, 
after the run is stopped, through a software based on custom-made C++ routines.

\section{Tests and results}

To test the performances of the planned SiRO detector, i.e.~its properties, as well as 
its fluctuations related to bias voltage, temperature, or the distance between the 
interaction point and the SiPM device, two detection modules have been built and placed in 
boxes for optical screening.

\subsection{The description of the SiPM devices}

To interpret the light signal produced at the interaction of a charged particle with 
the sensitive volume of the detector, devices like photomultipliers or photodiodes are 
used to convert the light yield into electrical output through photoelectric effect. 
Similar to photodiodes, a Silicon Photomultiplier is a semiconductor device, but 
their sensitive volume is divided into a matrix of hundreds of independent micro-cells, 
also named pixels, connected in parallel. Each micro-cell is operated in Geiger mode, 
the output signal of the SiPM device being proportional with the number of 
independent pixels triggered simultaneously.

A lot of interest is devoted to SiPM devices in the last years ~\cite{Retière, Alvarez, Sahin, Aguilar}. 
The properties of SiPM devices, like gain, after pulsing, cross-talk, dark noise, 
were carefully investigated in various studies~\cite{Eckert,Abe,Eraerds, Dolinsky, Vinogradov}.  

\begin{figure}[h]
\begin{center}
\includegraphics[width=\textwidth]{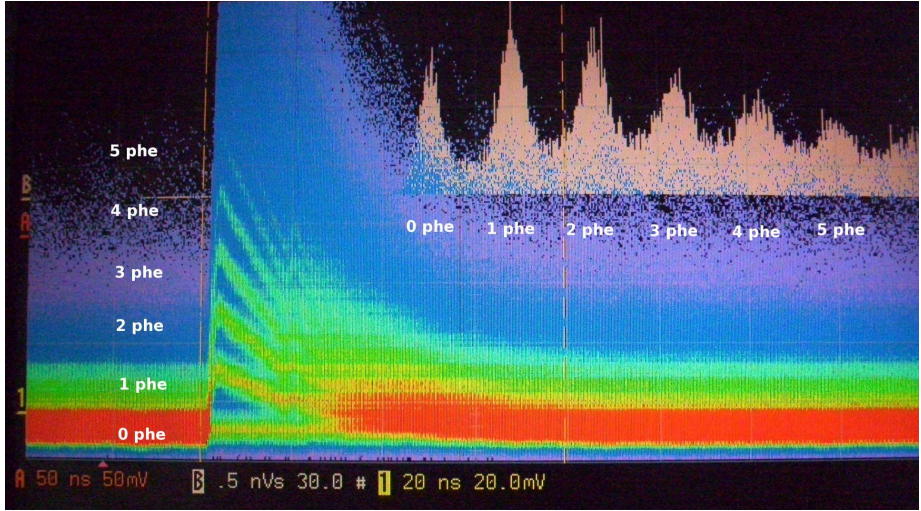}
\caption{\label{fig4} 
Oscilloscope view of a laser triggered MPPC S10362-33-100C~\cite{Stanca_1}.}
\end{center}
\end{figure}

We choose for our detector the MPPC S10362-33-100C model from Hamamatsu \cite{Hamamatsu}, 
with an effective area of $3~\times~3\,$mm$^2$. This device have been tested particularly at the 
Max-Planck-Institute of Physics in Munich~\cite{Stanca_1} with an experimental setup 
consisting of a stable voltage source, a signal amplifier (factor 50), a pulsed laser 
(pulsed diode laser PDL 800-B from PicoQuant) controlled by a signal generator 
(Synthesized Function Generator – model DS345 - from Stanford Research Systems). 
The output signals have been observed using a LC684DXL 1.5 GHz oscilloscope. 
Using the laser to trigger the events, we observed (Fig.~\ref{fig4}) the peaks 
corresponding to one to five photoelectrons, as well as the thermal noise.

We also found that the device is very sensitive to the variation of the bias voltage: 
an increase in gain with increasing voltage, but, also an proportional increase of unwanted 
effects with the gain, such as cross-talk and after pulsing effects. 
A significant sensitivity with the temperature was observed as well. 
Another aspect is that there are important differences from one device to another, 
so in order to have a uniform response, an individual analysis and calibration is required. 
The dependence of the gain on the operating conditions of the SiPM devices, like bias voltage and 
operating temperature, were observed in other studies as well, see, e.g.~\cite{Li, Licciulli}.

\subsection{Preliminary tests}

The first step was to test the response of a Detection Channel (DC) to muons, 
as well as how this response is changed with the applied bias 
voltage~\cite{Stanca_1,Niculescu,Stanca_2}. 

Measurements have been performed with two scintillator test probes S1 and S2 of size
$10~\times~10~\times~5\,$cm$^3$, each connected to a photomultiplier, with the tested DC placed 
between them. In order to ensure that the registered incident particles are from the defined
solid angle spanned by S1 and S2, the coincidence between the three devices DC, S1 and S2, 
is made. 
After a particle interact with all three sensitive volumes, the resulting three signals 
are passed through a FRONT END \& TRIGGER module. First, the pulses are passed through 
a variable threshold voltage comparator, being transformed from analogical type into a 
logical one with variable length, given by the time that the input signals stayed 
over the threshold. After that, those pulses are passed through a monostable circuit, 
forming output signals constant in length (100 ns) and amplitude, then through the 
coincidence circuit. The resulting pulse is registered by a SCALER-TIMER module.

\begin{figure}[h]
\begin{center}
\includegraphics[width=\textwidth]{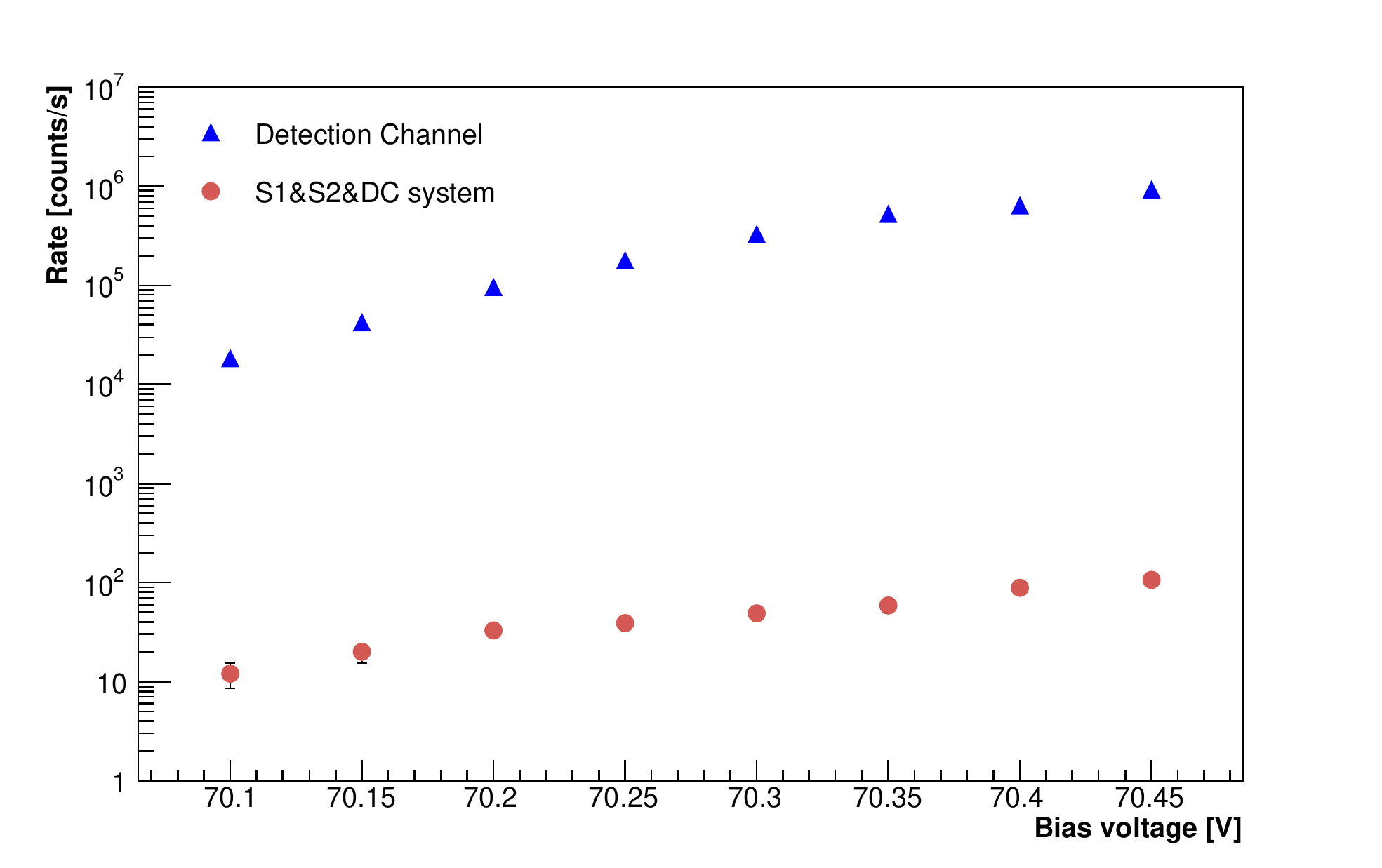}
\caption{\label{fig5} The variation of rates with the bias voltage of the SiPM~\cite{Stanca_2}.}
\end{center}
\end{figure}

We measured the variation of the pulse rate from the DC and the coincidence 
rate of S1~\&~S2~\&~DC with the bias voltage of the SiPM device, the temperature 
being kept constant and the bias voltage of S1 and S2 also constant. The results are displayed 
in Fig.~\ref{fig5}. It is easy to see that both, the S1~\&~S2~\&~DC coincidence rate and the 
DC rate are increasing with the increase of the bias voltage.

Taking into account that pulses from the SiPM device have about 200 ns in duration, 
for limiting the pileup effect, the average period of random pulses obtained should be 
ten times bigger than the pulse duration, which is $2\,\mu$s in our case and which 
corresponds to a maximum allowed rate of $500.000\,$counts/s. 
An optimum bias voltage has been chosen at $70.35\,$V, so the DC rate to be small 
enough and the S1~\&~S2~\&~DC rate to be as high as possible.

By measuring the coincidence rate between superposed S1 and S2 test probes, a rate of 
167 counts/1000s has been obtained. This means that for DC to have 100\% relative efficiency 
with respect to S1\&S2, the coincidence rate between S1~\&~S2~\&~DC needs to be 167 counts/1000s. 
Taking into account the optimum bias voltage that we choose, the relative efficiency of the 
DC as compared to S1 \& S2 coincidence rate is $\approx 35$\%.

\subsection{Investigation of the light response}

The light attenuation in respect to distance between the interaction point of muons 
with the sensitive volume and the SiPM device position is an important feature for the 
readout of scintillator detectors through Wave-Length-Shifter fibers. 
Such investigations were aleady reported by different 
experiments~\cite{Mineev,Vacheret,Mitrica}.

Tests have been performed upon a detection channel module (DC) to find 
how the light response is attenuated~\cite{Niculescu,Stanca_2}. We used for this purpose a 
LED diode, with the emission wavelength in the blue spectrum, steered by a 
signal generator (AFG TEKTONIX 3252). 

\begin{figure}[!h]
\begin{center}
  \begin{tabular}{cc}


    \includegraphics[width=60mm]{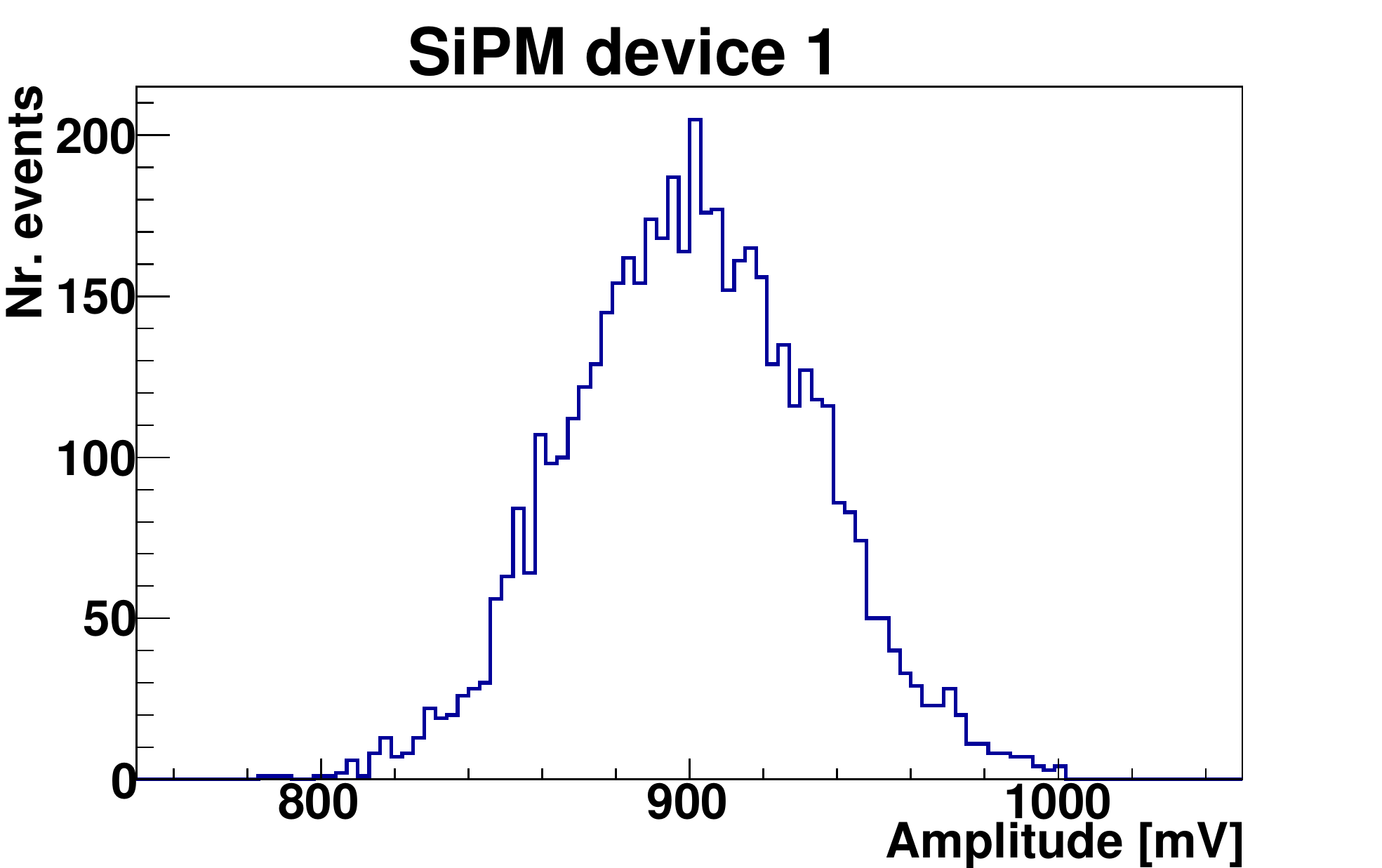}&

    \includegraphics[width=60mm]{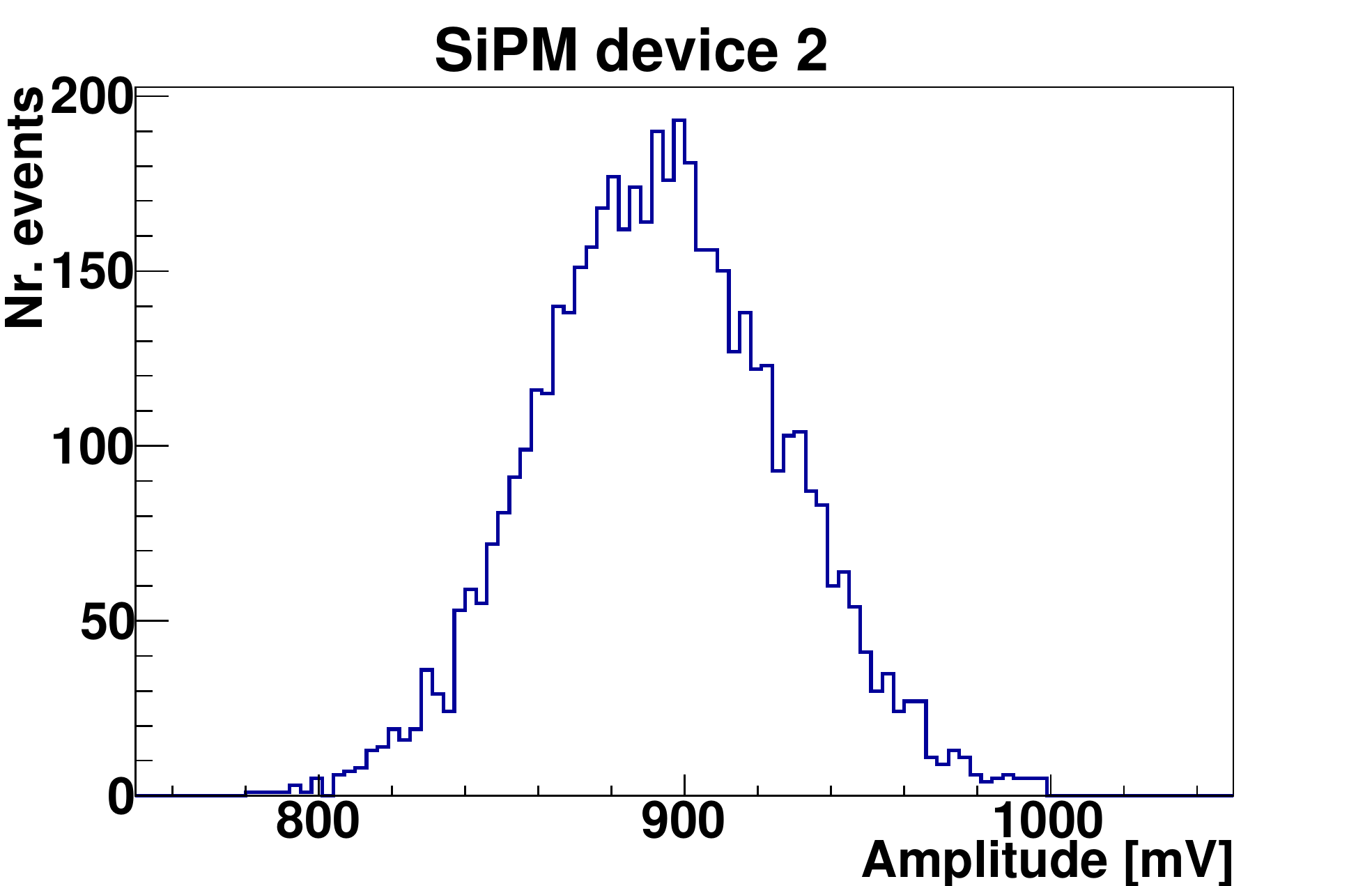}\\

    \includegraphics[width=60mm]{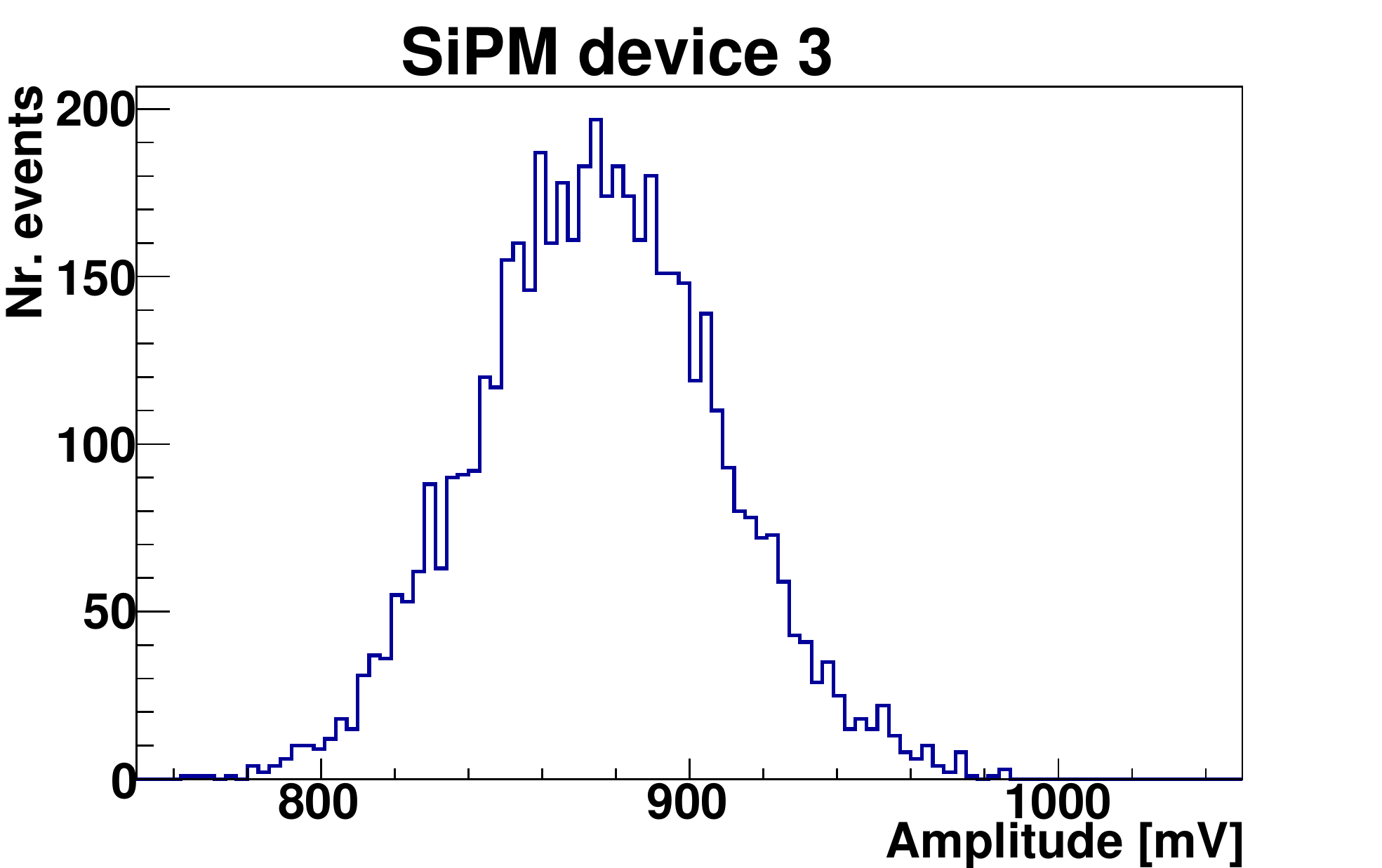}&

    \includegraphics[width=60mm]{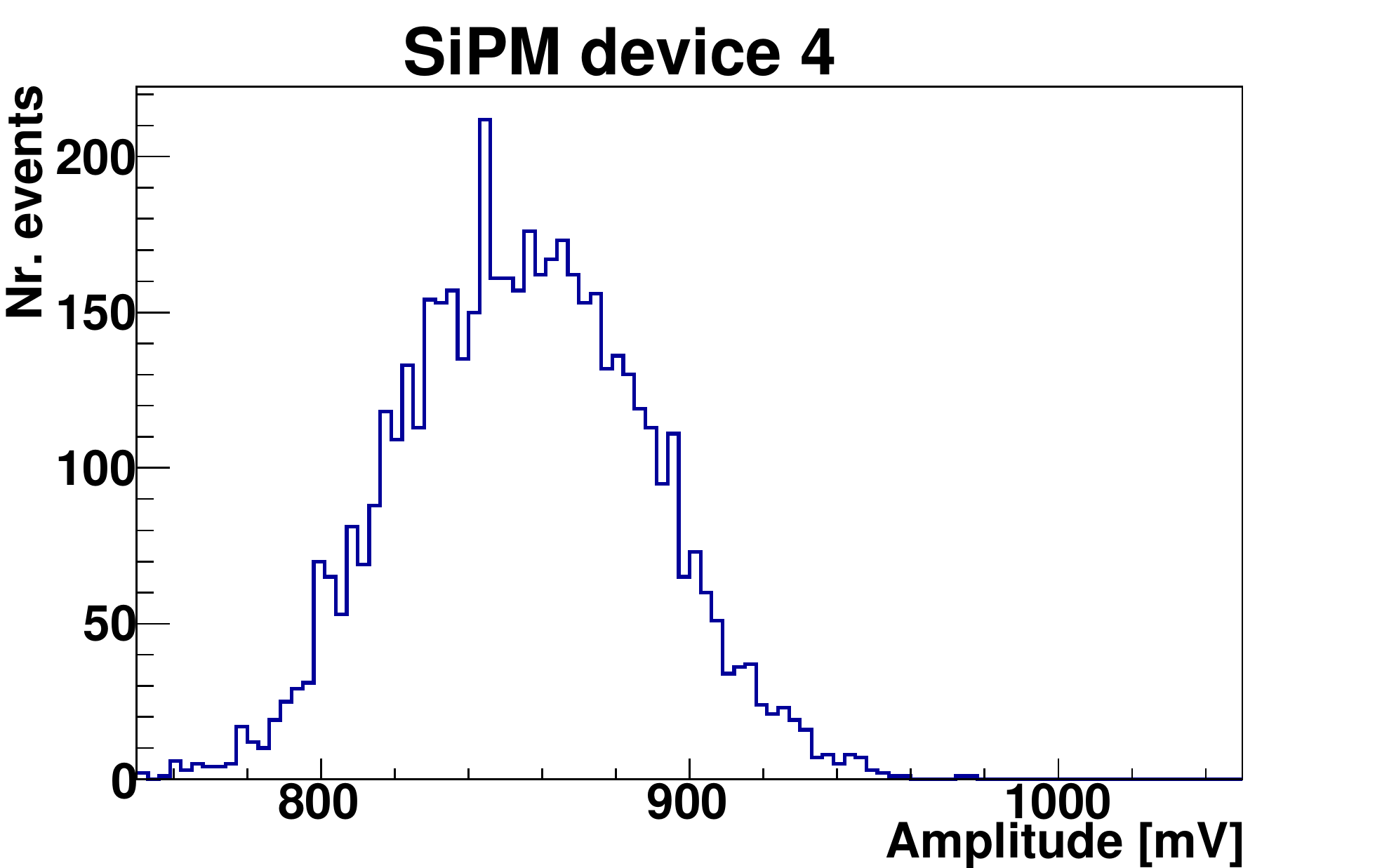}\\

    \includegraphics[width=60mm]{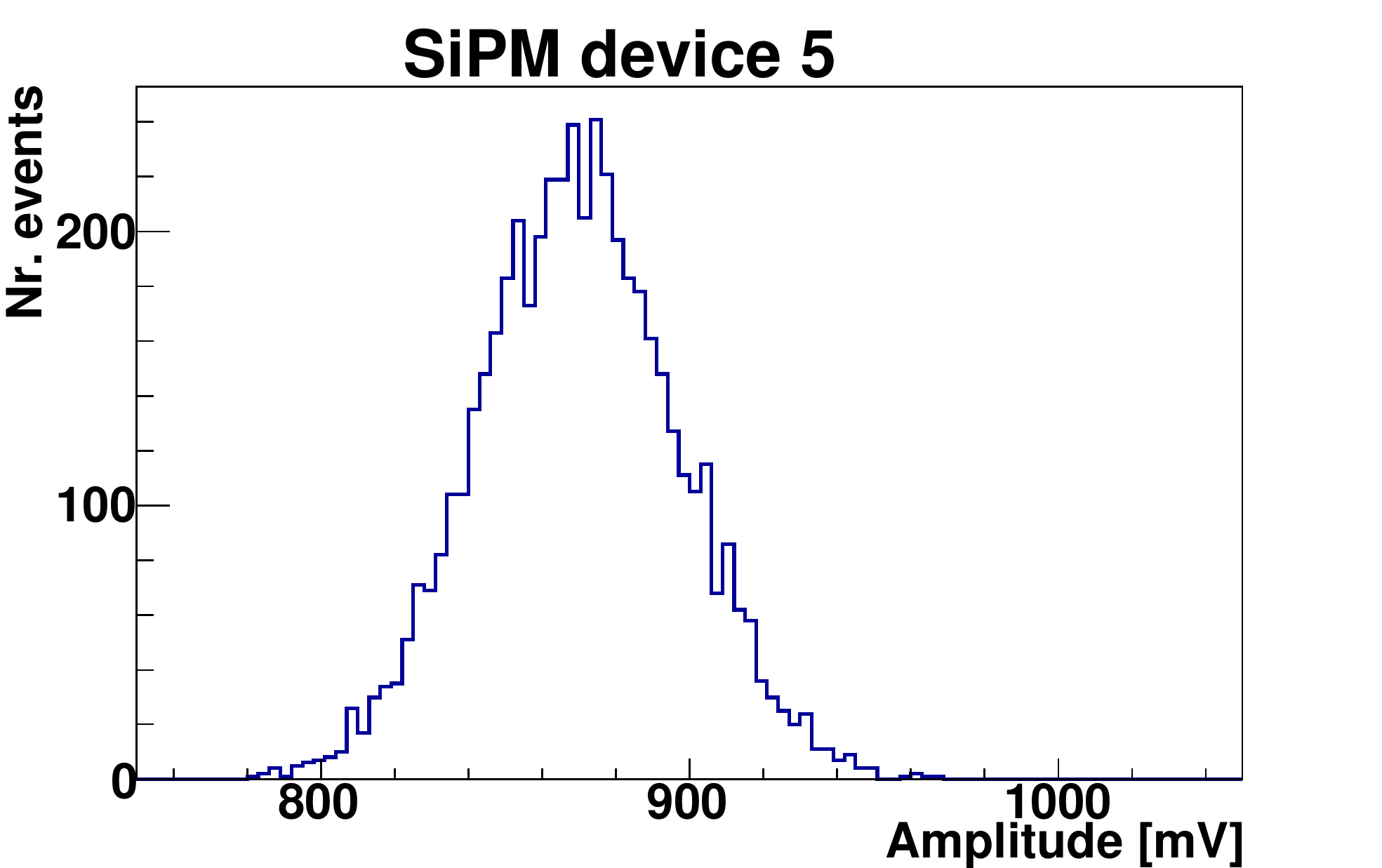}&

    \includegraphics[width=60mm]{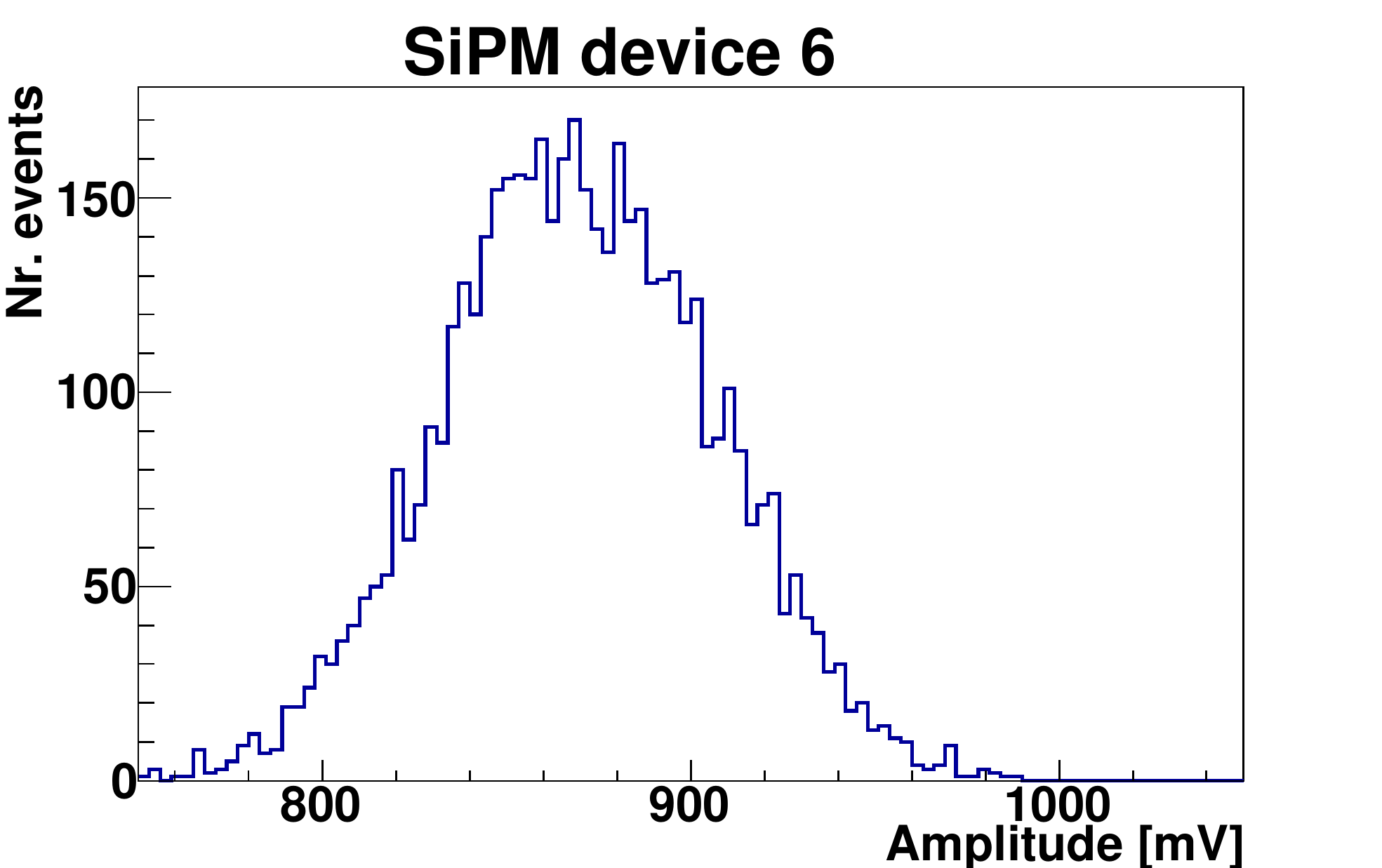}\\

  \end{tabular}
\caption{\label{fig6} Examples of the signal amplitudes distributions recorded 
for each channel that compose one detection module after the excitation with an 
LED diode by a 64 channel digitizer (V1740 CAEN) \cite{Niculescu}.}
\end{center}
\end{figure}

First, the responses of the channels have been aligned by stimulating each optical fiber 
with the LED device and setting the SiPM devices output signal to a fixed value 
through bias voltage corrections. The distribution of the signal amplitudes given 
by each SiPM is shown in Figure~\ref{fig6}. The LED excitation was fixed in a way that 
the maximum amplitude of the signal on every channel to be in the range of $0.8\,$V - $0.9\,$V.

Different attenuation tests have been performed: 

{\it i) The signal attenuation across the 6 channels that compose a detection module}

\begin{figure}[!h]
\begin{center}
\includegraphics[width=\textwidth]{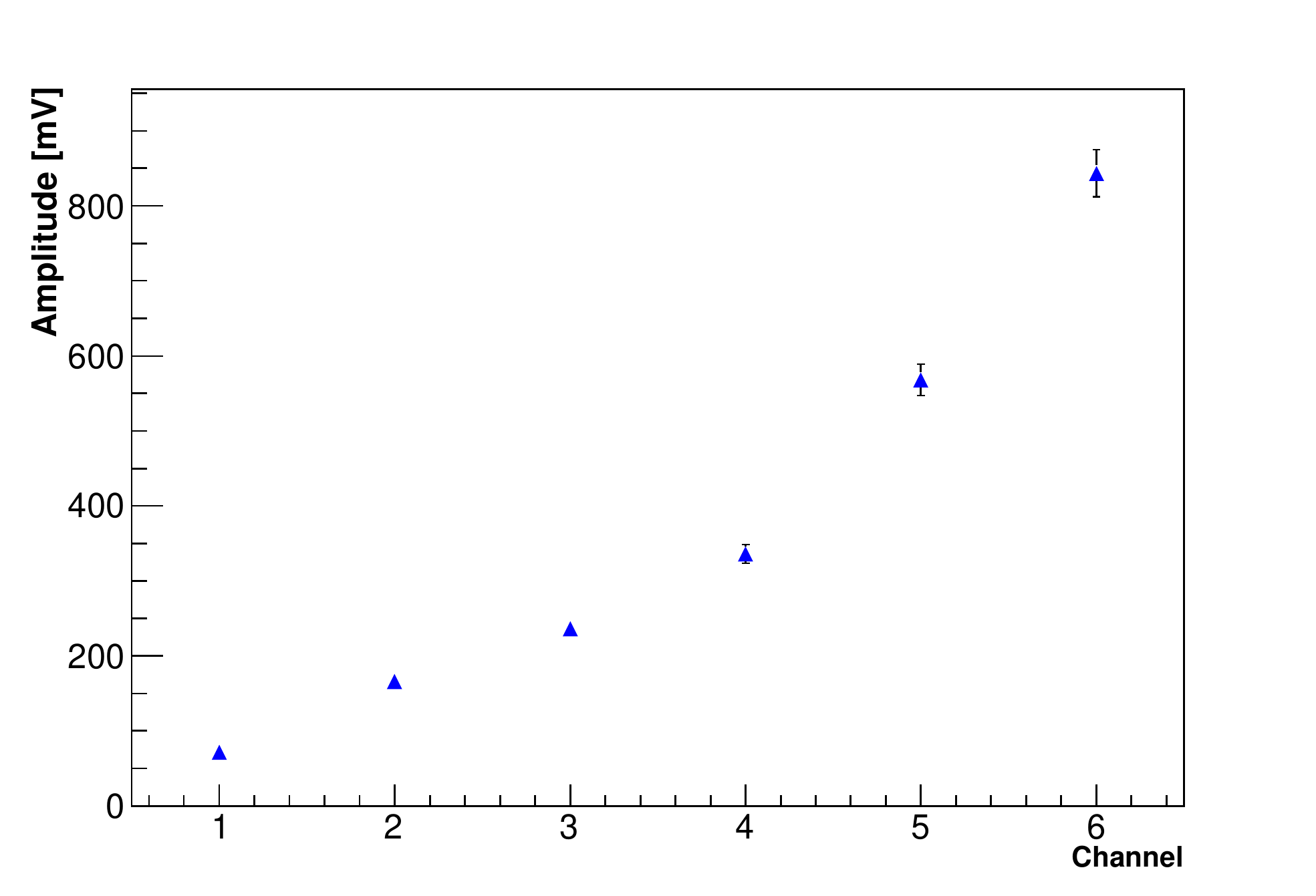}
\caption{\label{fig7} The amplitude of the signals registered by all six channels that 
form a detection module when the module is triggered by a LED diode in the vicinity 
of the sixth channel~\cite{Niculescu}.}
\end{center}
\end{figure}

The responses of all six channels were registered during the stimulation of the 
6th channel with the LED diode at 96 cm distance from the corresponding SiPM device. 
As we can see in Fig.~\ref{fig7}, a 33 \% attenuation on the neighboring channel was 
obtained (which is not enough for a clear amplitude discrimination), 
followed by a more abrupt decrease of the signals amplitudes in the remaining channels.

{\it ii) The signal attenuation along one single channel}

\begin{figure}[!h]
\begin{center}
\includegraphics[width=\textwidth]{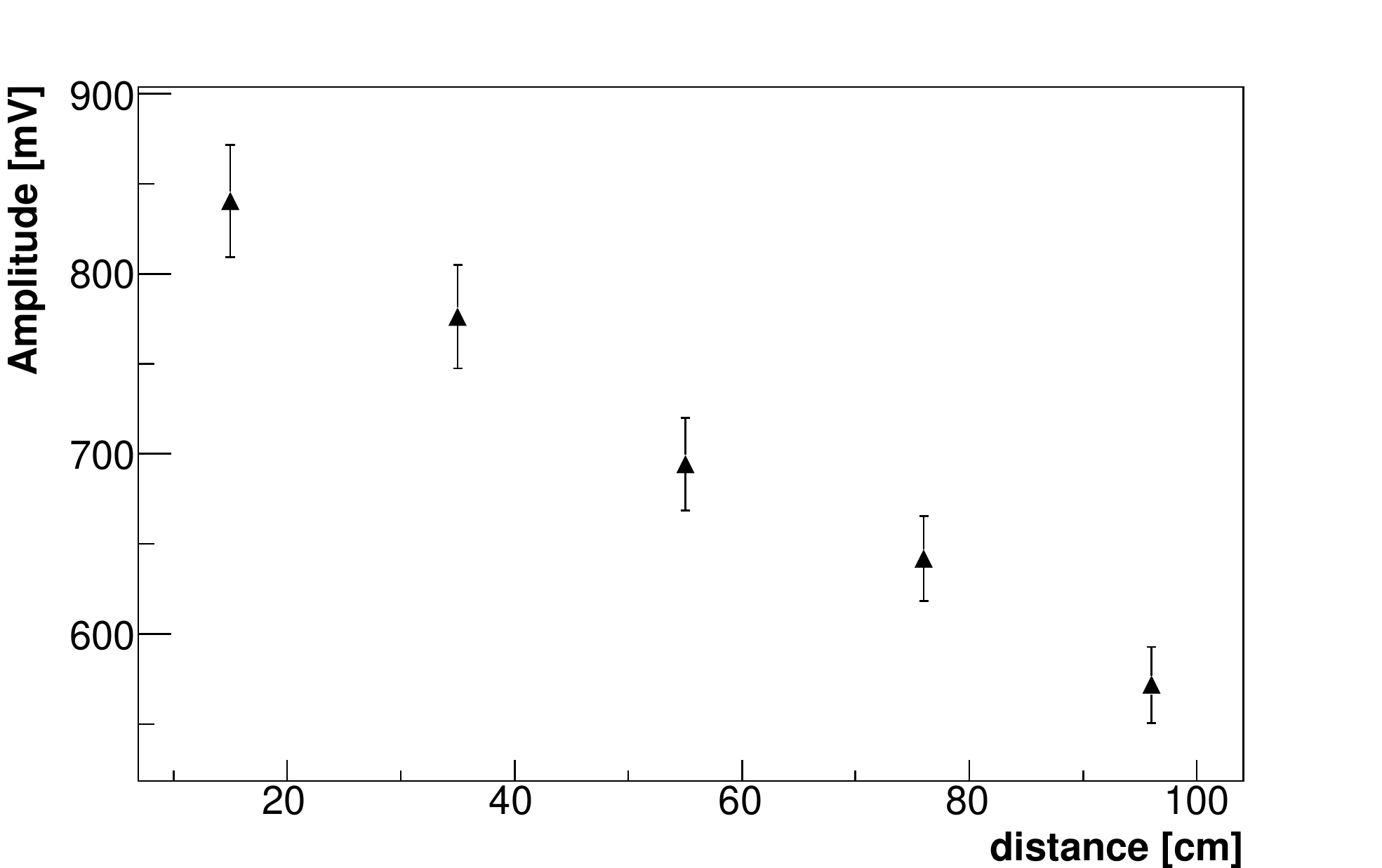}
\caption{\label{fig8} The signal amplitude attenuation versus the distance between the 
interaction point and the SiPM device~\cite{Brancus, Stanca_2}.}
\end{center}
\end{figure}

Measuring the amplitudes of the signals on the same detection channel when the LED 
excitation is applied at different distances along one of the channel’s optical fibers, 
the light attenuation curve shown in Fig.~\ref{fig8} is obtained. 
The drop in light response is significant, a 43 \% light attenuation is observed at positions 
between 17 and 97 cm. 

\subsection{Testing the detection module response at different temperature gradients}

To measure how temperature affects our results, one detection module has been tested 
underground in the Unirea salt mine at Slanic Prahova, Romania, at different ambient 
temperature in the interval 13$^o$ to 20$^o$ Celsius. The bias voltage for each channel 
was selected that, at 13$^o$ Celsius, its individual counting rate to be at about 400 counts/s.

\begin{figure}[!h]
\begin{center}
\includegraphics[width=\textwidth]{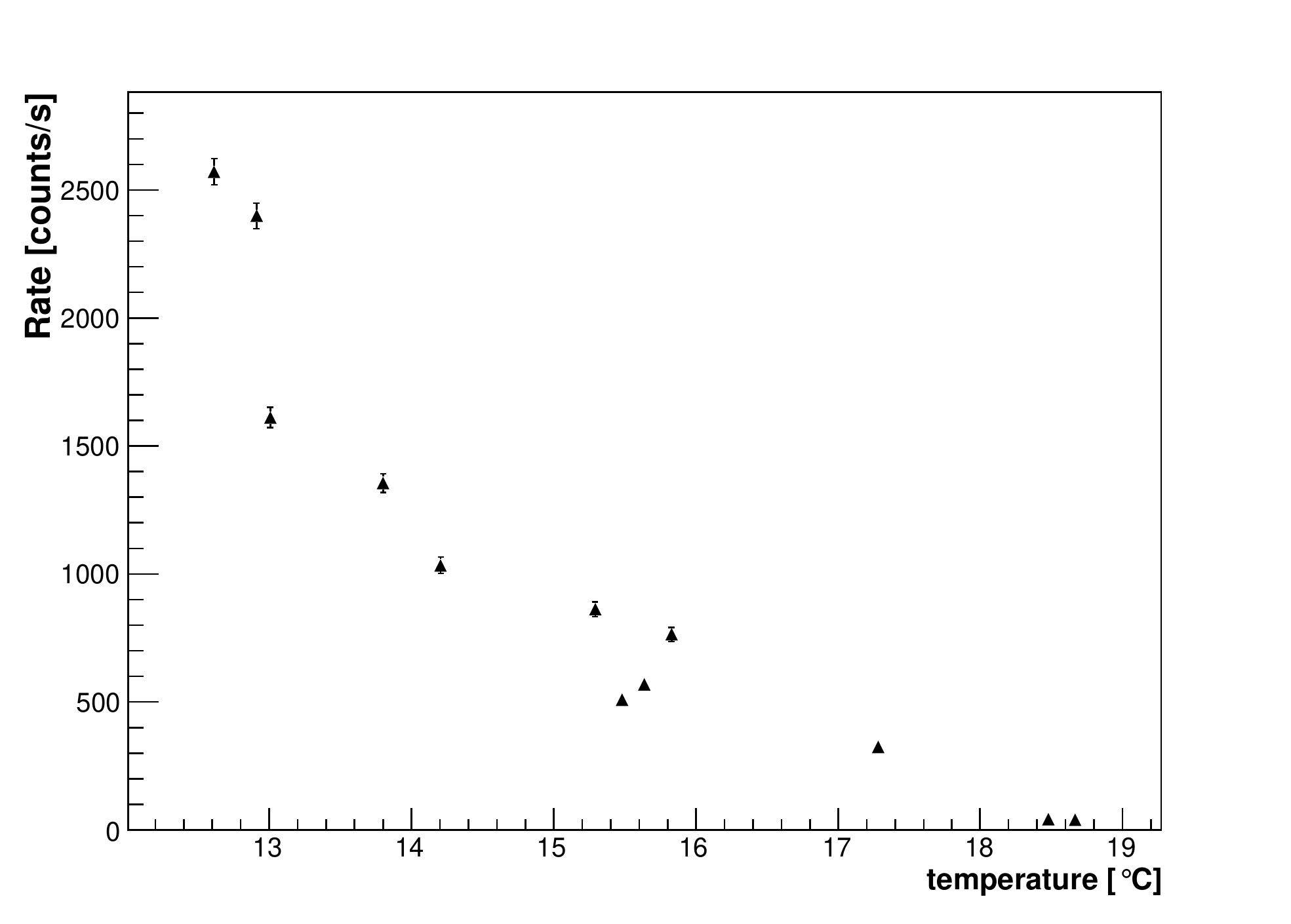}
\caption{\label{fig9} The detection module response, obtained by summing up the output of 
all six channels, versus the environmental temperature.}
\end{center}
\end{figure}

The results plotted in Fig.~\ref{fig9} shows that the rates are dramatically decreasing 
with the increase of temperature. A fluctuation of 1$^o$ C will change significantly 
the response of the SiPM devices. Taking this into account, we conclude that the final 
configuration of the SiPM muon detector should be placed in an environment with high 
thermal stability, like the underground Unirea salt mine from Slanic Prahova, where 
the environmental temperature is about 12$^o$C-13$^o$ C in every moment of the day or 
the year.

\subsection{Tests of muon detection using a two detection modules system}

\begin{figure}[!h]
\begin{center}
\includegraphics[width=\textwidth]{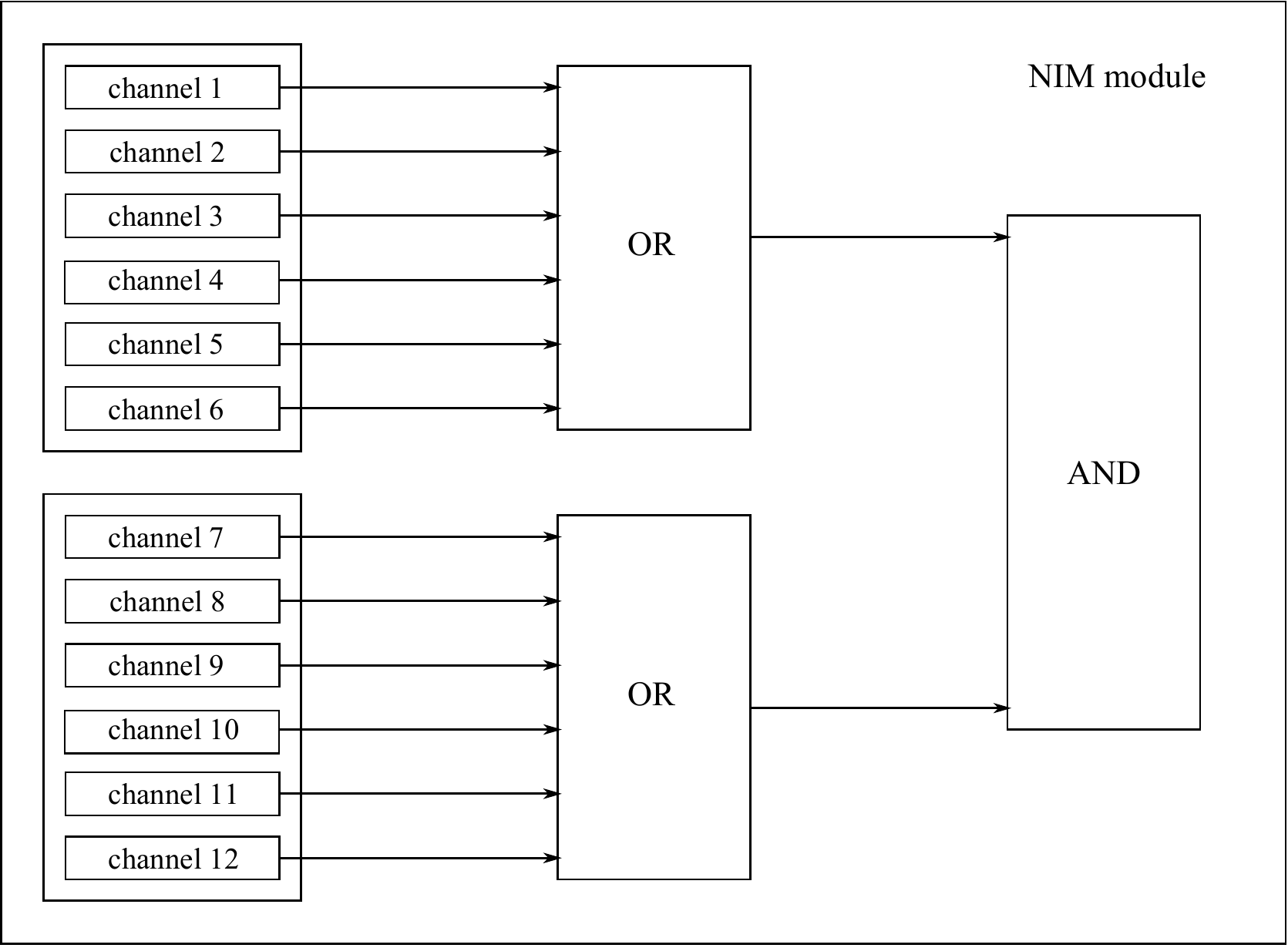}
\caption{\label{fig10} The block diagram of the NIM module.}
\end{center}
\end{figure}

Two detection modules, named D1 and D2, were put in coincidence to test their response 
to muons. 
For this, a 12 channels NIM module was used. Each NIM channel is equipped with a 
variable threshold comparator. The first 6 channels corresponds to D1 and are 
summarized by an OR function. A similar treatment is applied to the next 6 channels 
that corresponds to D2. Both OR outputs are put in coincidence by an AND function. 
The block diagram of the NIM module is represented in Fig.~\ref{fig10}. 

In this way, when a muon cross both detection modules, a coincidence signal is formed, 
which is used as trigger by a digitizer (DT5740 from CAEN) to transfer the recorded data 
to a PC for storage.

\begin{figure}[!h]
\begin{center}
\includegraphics[width=\textwidth]{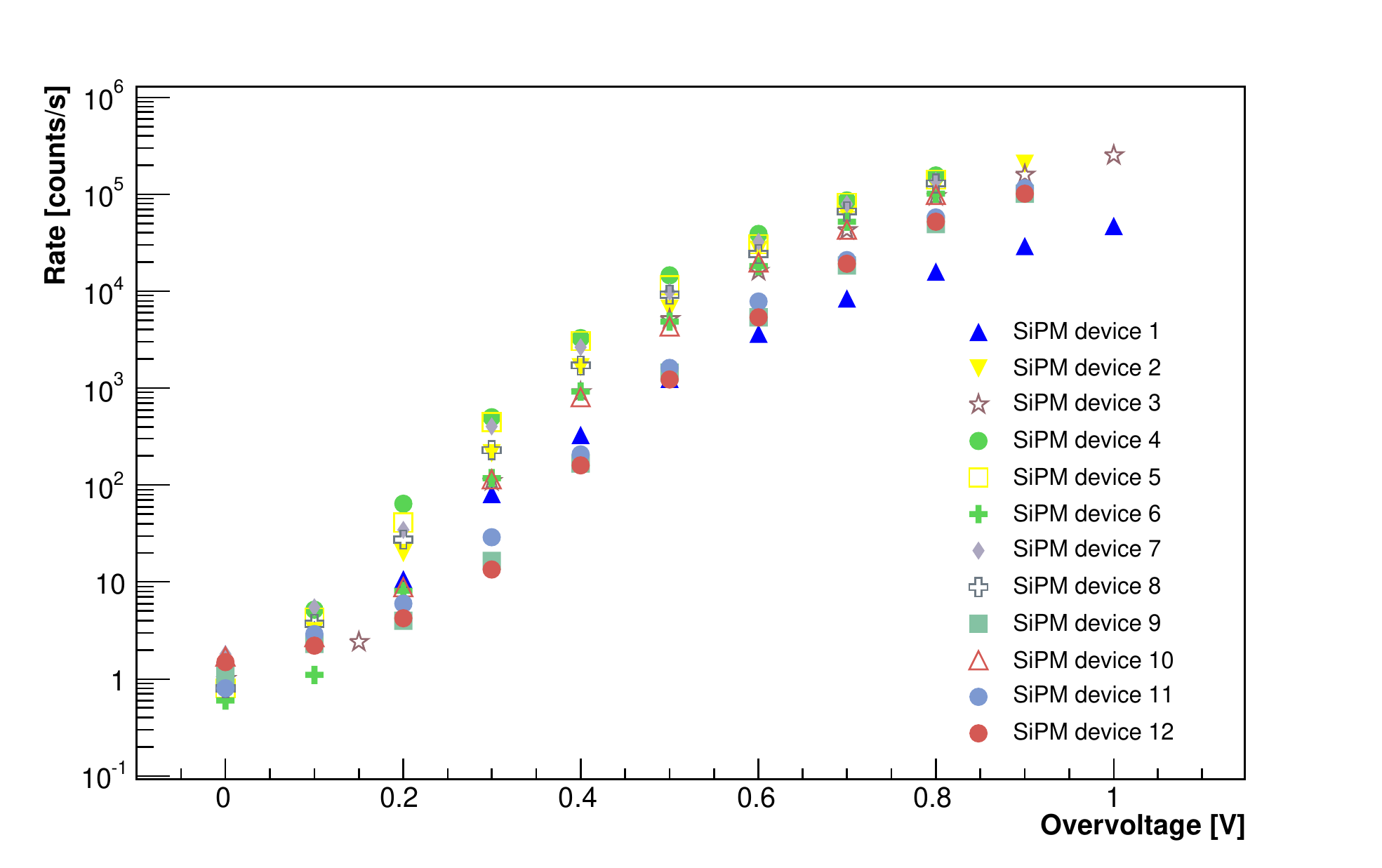}
\caption{\label{fig11} The counting rate dependencies on the 
difference between the applied bias voltage and the breakdown voltage, also 
known as overvoltage, for each SiPM device that are part of D1 and D2~\cite{Stanca_2}.}
\end{center}
\end{figure}

Before starting the measurements, each SiPM device has been tested. Fig.~\ref{fig11} shows 
the dependence of the SiPM counting rates on the applied bias voltage. 

\begin{figure}[!h]
\begin{center}
  \begin{tabular}{cc}


    \includegraphics[width=60mm]{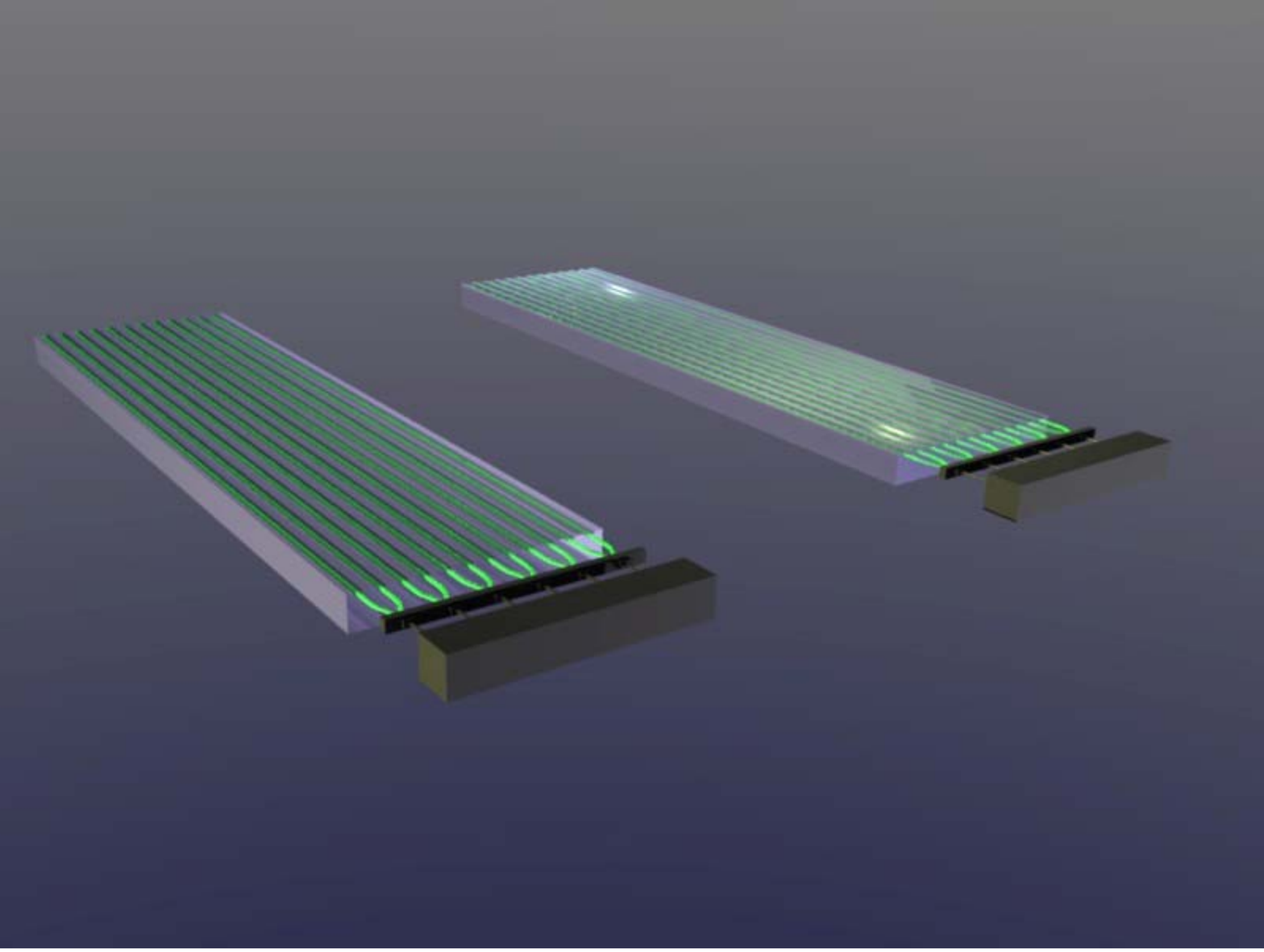}&

    \includegraphics[width=60mm]{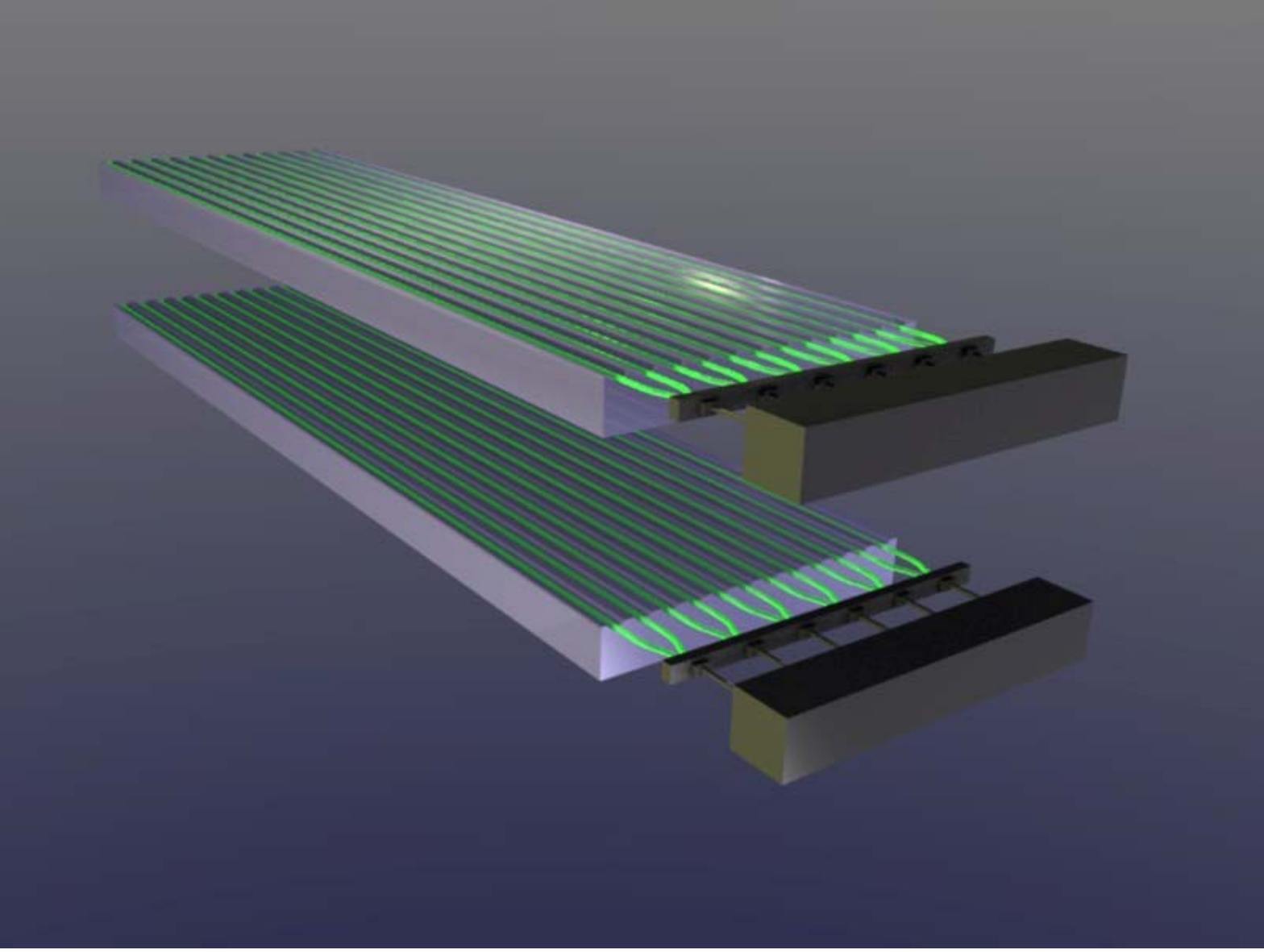}\\
a) & b)\\

    \includegraphics[width=60mm]{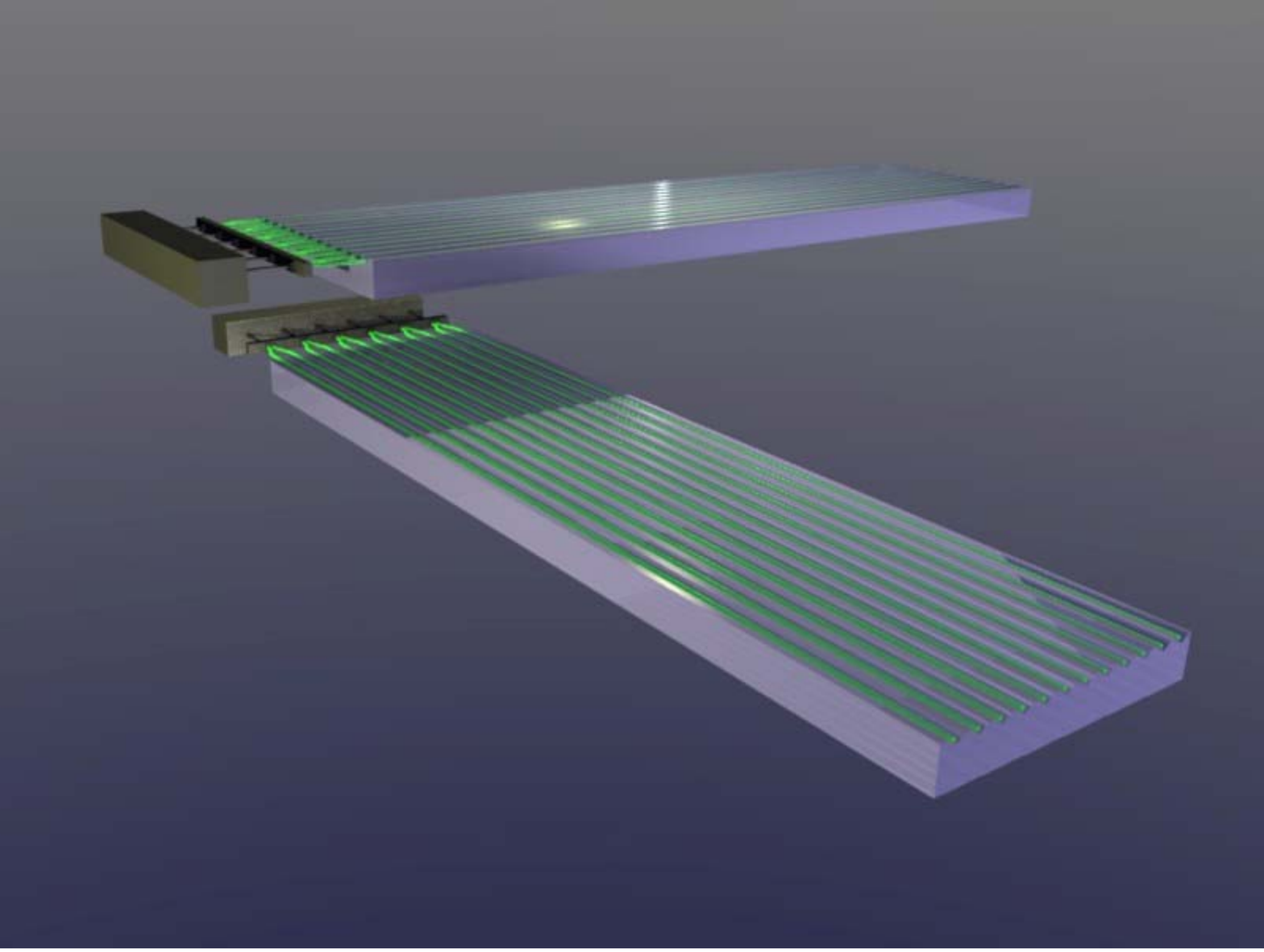}&

    \includegraphics[width=60mm]{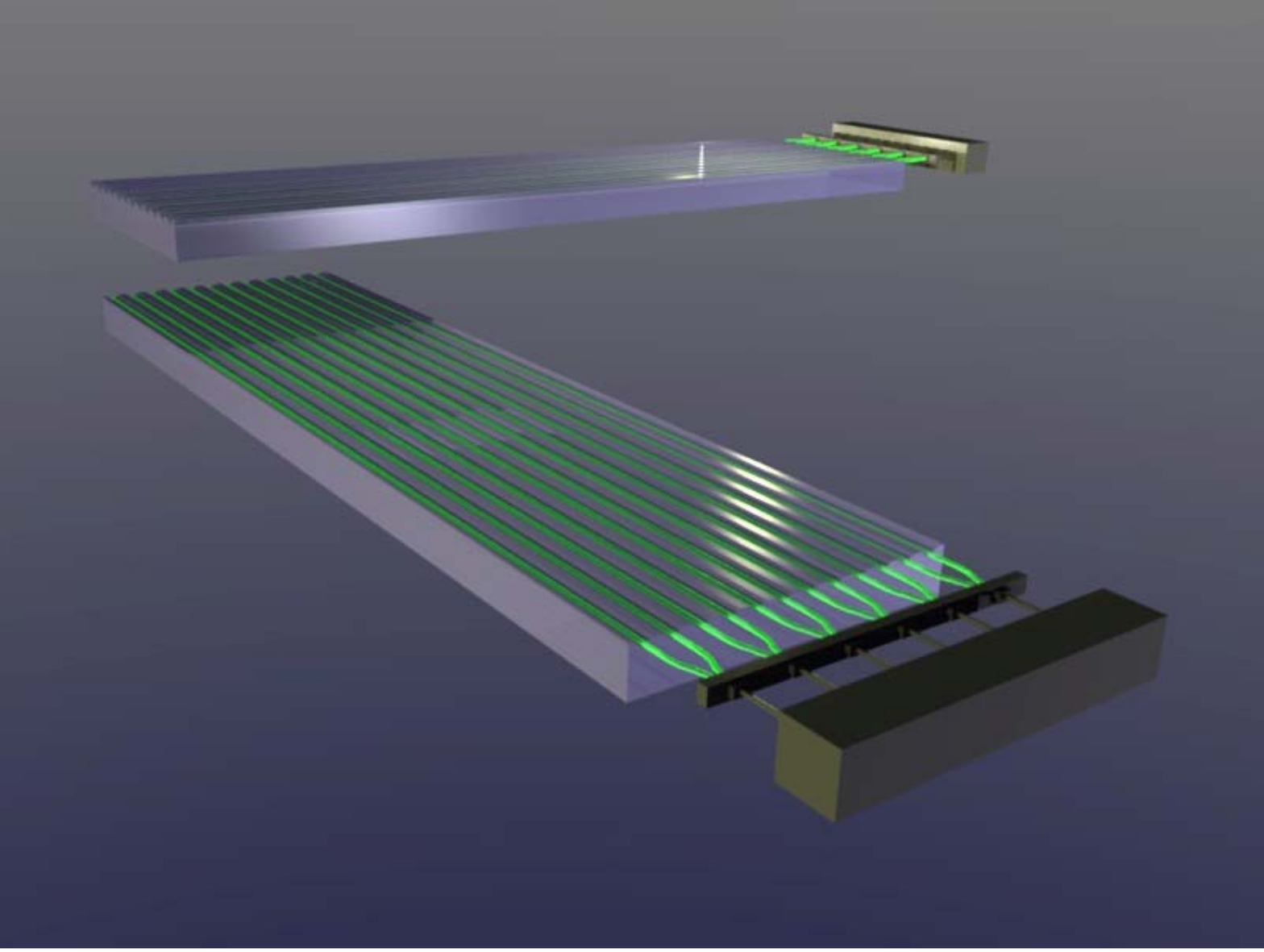}\\
c) & d)

  \end{tabular}
\caption{\label{fig12} Different configuration of the subassembly D1 and D2~\cite{Stanca_2}; 
a) D1 and D2 placed on a plane surface with 70 cm in between; 
b) D1 and D2 totally superposed; 
c) D1 and D2 superposed at the collection side; 
d) D1 and D2 superposed at the opposite side of collection. }
\end{center}
\end{figure}

The measurements have been made using with D1 and D2 detection modules placed in 
different configurations, as it is shown in Fig.~\ref{fig12}, with the individual SiPM 
bias voltages set for the detection modules responses to fit in four different rate 
intervals (R1 to R4). The results are presented in Table~\ref{tab1}.

 It is observed an  increase of the coincidence rate obtained with the D1 and D2 system 
with the increasing of the individual SiPM rates due to false coincidence events. Increasing 
the rate from the R1 rate interval to that of R4, leads to a gain increase 
of the response of the SiPM devices, but also to a much steeper increase of the crosstalk 
effect. It can also be seen that the coincidence rate between D1 and D2 is strongly 
dependent on their placement within the 4 configurations. 
Taking that into account, the R1 rate interval was chosen for further testing. The 
configuration (b), with the modules D1 and D2 totally superposed, is being used.

\begin{table}[!h]
\begin{center}
\caption {\label{tab1} Coincidence measurements with D1 and D2 placed in 
different configurations, for different individual SiPM rate intervals }
\vspace*{0.5cm}
\begin{tabular}{|c|c|c|c|c|c|c|}
\hline
 Rate& D1 rate & D2 rate & D1 and D2 & D1 and D2 & D1 and D2 & D1 and D2 \\
 intervals& $[counts/s]$ & $[counts/s]$ & $[counts/s]$ & $[counts/s]$ & $[counts/s]$ & $[counts/s]$ \\
  & & & (a) & (b) & (c) & (d) \\
\hline

R1	& 712            & 722           & 0,25           & 10,98         & 5,17          & 2,24 \\
\hline
R2	& 3126	& 3345	& 2,05	& 16,78	& 8,14	& 4,69 \\
\hline
R3	& 7110	& 7232	& 8,15	& 21,13	& 9,79	& 8,67 \\
\hline
R4	& 62.702      & 70.437       & 734   	& 763	          & 745	          & 744 \\
\hline
\end{tabular} 
\end{center}
\end{table}

\subsection{Testing amplitude pulse discrimination}

The aim of those measurements were to investigate, how well we can determine (by tracking 
the position) where the incident muons interact with the sensitive volumes of D1 and D2 
modules based on an amplitude pulse discrimination.

Coincidence measurements were performed with the D1-D2 detection system described above 
where the individual rates for each SiPM device was fixed through bias-voltage manipulation 
for the detection modules responses to fit an interval rate of 700-800 pulses/s. 

A C++ routine was developed to analyze the stored data~\cite{Niculescu} for determining 
if a registered event, triggered by the NIM module, is a valid one and can be used to 
determine the muon trajectory. 

The routine consists of two parts:
In the first part, the peak amplitudes for all channels are counted, the peak distribution 
are obtained, and  the average value for each channel response is calculated.
The second part analyzes each event triggered by the digitizer and decide if it can be 
accepted as a valid one.

For an event to be valid, the following conditions must be fulfilled:
The peak amplitude should be at least 70 mV high. The time difference between the peak 
generated in the top module and in the bottom module should be lower than 32 ns. A 
minimum of 2 peaks on adjacent channels have to be available at both top and bottom module.

\begin{figure}[h]
\begin{center}
\includegraphics[width=\textwidth]{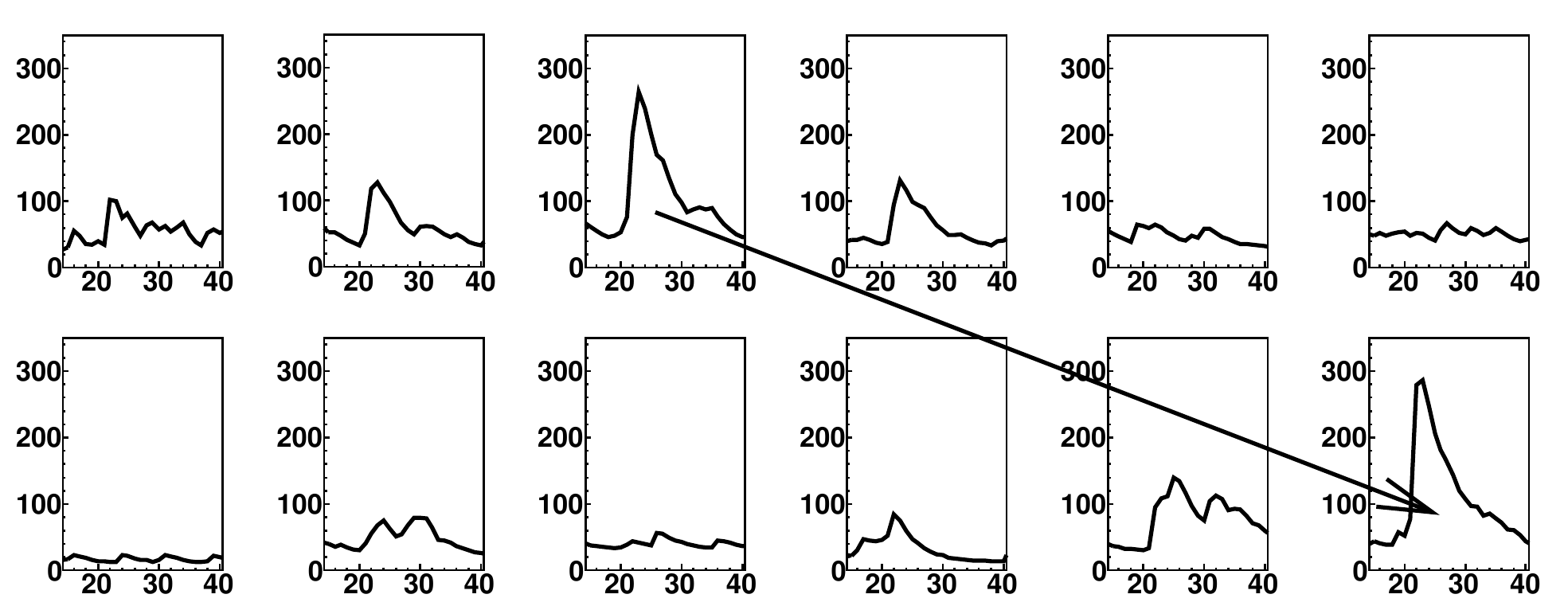}
\caption{\label{fig13} Example of a valid event. The top spectrum represents D1 channels, 
the bottom one represents the channels from D2~\cite{Niculescu}.}
\end{center}
\end{figure}

In Fig.~\ref{fig13}, a valid event is exemplified where the trajectory is determined by 
software. When such a trajectory cannot be found we flag the event as invalid.

The efficiency of the D1-D2 detection system to determine the trajectory of the incident 
muons was obtained by analyzing a number of 56749 events triggered by the NIM Module in 
7 different runs. The software routine validated a number of 17169 events, 
resulting in a 29.1\% efficiency value.

\section{Concluding remarks and outlook}

The properties of the MPPC S10362-33-100C device from Hamamatsu have been investigated 
(attached to fibers grooved in bars of scintillation detectors),
especially their behavior in respect to bias voltage changes and temperature fluctuations.

Tests have been performed with a detection module of the detector to define its characteristics.
After calibrating each channel with a LED device, the attenuation of the light signal 
has been investigated along the fibers that forms a channel and across the channels 
that form the detection module.

Using two superposed detection modules, placed in different configurations, the muon 
response was tested for different SiPM rate intervals.
Comparing the results obtained with the two detection modules placed in a parallel 
side by side configuration or placed parallel one above the other, it is observed that there 
are small differences when the SiPM devices are set to fit to the 7000 counts/s 
and the 70 000 counts/s rate intervals. 
This is mainly due to high crosstalk and dark noise rates of the SiPM devices, 
leading to the conclusion that those individual rate intervals are unreliable.

Therefore, the parallel configuration with the detection modules placed one above the 
other was chosen for further testing, with the SiPM individual rates being fixed to a total 
rate of the detection modules to fit the interval of 700-800 counts/s.

The possibility to determine the incidence point in the input \textit{XY} plane of the 
incident muon using two detection modules through the amplitude pulse discrimination 
method has been proven to be valid. Unfortunately, we find out that the efficiency 
of this method is only about 30 \%, i.e.~not enough to continue using it as a basic 
concept for the design of the planned multi-purpose, mobile SiRO detector.  

	Based on the tests and results presented in this paper, we decided to develop a new method, 
by replacing the old detection module that sum-up all 6 channel responses by 6 
individual channels obtained by slicing its $25~\times~100~\times~1\,$cm$^3$ sensitive volume 
into $4~\times~10~\times~\,$cm$^3$ stripes and replacing the analog acquisition technique 
by a digital one. 
By this it is expected to improve the performance and the stability of SiRO, including a better 
use of the aforementioned advantages of the readout based on SiPMs. \\

{\bf Acknowledgments}

This work is supported by the Romanian Authority for Scientific Research UEFISCDI, 
PNII-IDEI grants 17/2011 and PN 09 37 01 05. A.~Badescu would like to thank the 
Sectoral Operational Programme Human Resources Development 2007-2013 of the Ministry 
of European Funds for the Financial Agreement POSDRU/159/1.5/S/132395. We are grateful 
to Prof. M. Teshima and Prof. R. Mirzoyan for fruitful discussions and for the kind 
hospitality in performing the tests with SiPM at Max-Planck-Institut for Physics. 
We acknowledge the essential contribution of DE Mirel Petcu for the design and the setting up of 
the SiRO detector.
We thank to A.-M.~Crisan for his contribution regarding the electronic part of the 
SiPM muon detector. 



\end{document}